\documentclass[a4paper,11pt]{article}
\pdfoutput=1 

\usepackage{jheppub} 

\usepackage[T1]{fontenc} 

\newcommand{\nb}{\nonumber}
\def\be{\begin{equation}}
\def\ee{\end{equation}}
\newcommand{\bea}{\begin{eqnarray}}
\newcommand{\eea}{\end{eqnarray}}
\def\bfig{\begin{figure}}
\def\efig{\end{figure}}

\newcommand{\ud}{\mathrm{d}}

\def\dd{{\rm d}}
\def\m{\mu}
\def\n{\nu}
\def\a{\alpha}
\def\b{\beta}
\def\g{\gamma}
\def\l{\lambda}
\def\s{\sigma}
\def\o{\omega}
\def\de{\partial}
\def\j{\varphi}
\def\De{\nabla}
\def\xx{{\bf x}}
\def\V{{\cal V}}
\newcommand\Ostro{Ostrogradski }

\title{\boldmath  Degenerate Ho\v rava gravity}

\author{Enrico Barausse,}
\author{Marco Crisostomi,}
\author{Stefano Liberati}
\author{and Lotte ter Haar}
\affiliation{SISSA, Via Bonomea 265, 34136 Trieste, Italy and INFN Sezione di Trieste\\IFPU - Institute for Fundamental Physics of the Universe, Via Beirut 2, 34014 Trieste, Italy}

\abstract{
Ho\v rava gravity breaks Lorentz symmetry by introducing a dynamical timelike scalar field (the khronon), which can be used as a preferred time coordinate (thus selecting a preferred space-time foliation).  Adopting the khronon as the time coordinate, the theory is invariant only under time reparametrizations and spatial diffeomorphisms. In the infrared limit, this theory is sometimes referred to as  khronometric theory.
Here, we explicitly construct a generalization of khronometric theory, which avoids the propagation of Ostrogradski modes as a result of a suitable degeneracy condition (although stability of the latter under radiative corrections remains
an open question).
While this new theory does not have a general-relativistic limit and does not yield a Friedmann-Robertson-Walker-like cosmology on large scales, it still passes, for suitable choices of its coupling constants, local tests on Earth and in the solar system, as well as gravitational-wave tests. We also comment on the possible usefulness of this theory as a toy model of quantum gravity, as it could be completed in the ultraviolet into a ``degenerate Ho\v rava gravity'' theory that could be  perturbatively renormalizable without imposing any projectability condition.}

\begin{document} 
\maketitle
\flushbottom

\section{Introduction}
Ho\v rava gravity~\cite{Horava:2009uw} is a gravitational theory that 
is power-counting renormalizable in the ultraviolet (UV), at the expense of
giving up Lorentz symmetry. The theory, written in terms of 3+1 
 Arnowitt-Deser-Misner  (ADM) variables~\cite{PhysRev.116.1322}, is indeed invariant only under 
  foliation-preserving diffeomorphisms (FDiffs), i.e.~(monotonic) time reparametrizations and spatial diffeomorphisms,
and \textit{not} under full-fledged four-dimensional diffeomorphisms. The  action of the theory involves up to six spatial derivatives
of the ADM fields, but is only quadratic in the time derivatives (which only appear
via the extrinsic curvature, i.e.~via the time derivative of the spatial metric).
It is this anisotropic scaling between space and time derivatives that  ensures power-counting renormalizability. 

Performing a Stuckelberg transformation,  Ho\v rava gravity can be recast as a Lorentz-violating scalar-tensor theory~\cite{Jacobson:2013xta}. The scalar field, 
sometimes referred to as the khronon, is constrained by a Lagrange
multiplier to be timelike (i.e.~its gradient must be timelike),  and it thus plays the role of a preferred time 
 by selecting a preferred manifold slicing. Using this covariant formulation, as opposed to the  original 3+1 one (i.e.~the unitary gauge where the khronon is adopted as time coordinate), it becomes more clear why 
 the 3+1 action is built without any
 time derivatives of the lapse. In fact,
 to avoid \Ostro instabilities \cite{Ostrogradsky:1850fid} (or ``ghosts''),
 one may naively require that the covariant action be quadratic in 
 the unit-norm ``\ae ther'' vector field proportional to the khronon's gradient. This vector field turns out to be $\boldsymbol{u}=-N dt$ in the unitary gauge, with $N$ the lapse, and one may  naively try to obtain time derivatives of $N$ by introducing
 the acceleration $a^\nu=u^\mu \nabla_\mu u^\nu$. The latter, however,
 only includes spatial derivatives of $N$ because of the unit norm condition
 (which implies $a^\mu u_\mu=0$).

 The absence of time derivatives of the lapse poses a hurdle to proving 
 perturbative renormalizability (beyond power counting). 
 Calculations of the latter are technically involved and have so far only been performed in the unitary gauge~\cite{Barvinsky:2015kil,Barvinsky:2017zlx}, where the lapse satisfies an elliptic equation
 as a result of the absence of its time derivatives in the action~\cite{Blas:2010hb}. This leads to
 the ``instantaneous'' propagator $1/(k_i k^i)$. To overcome this problem, Ref.~\cite{Barvinsky:2015kil} proved  perturbative renormalizability
 in Ho\v rava gravity under the ``projectability condition'', i.e.~the assumption that the lapse is a function of time only. 
  The resulting theory, while considered in the first paper by  
  Ho\v rava~\cite{Horava:2009uw}, is 
disjoint from general (i.e~non-projectable)  Ho\v rava gravity~\cite{Blas:2010hb},
and is strongly coupled on flat space~\cite{Blas:2010hb,Blas:2009yd,Koyama:2009hc}.

It should be noticed, however, that including derivatives of the lapse (i.e.~second time derivatives of the khronon scalar field) does not lead automatically to \Ostro ghosts, if the Lagrangian is {\it degenerate}. This fact is very well known in the context of Lorentz-symmetric scalar tensor theories, where it led first to beyond-Horndeski theories \cite{Zumalacarregui:2013pma,Gleyzes:2014dya} and then to Degenerate Higher-Order Scalar-Tensor (DHOST) theories \cite{Langlois:2015cwa, Crisostomi:2016czh, BenAchour:2016fzp}. These theories have field equations that are higher than second order in time, but still propagate no ghosts. 

In the following, we will apply this degeneracy program to the infrared (IR) limit of non-projectable (i.e.~general) Ho\v rava gravity. In that limit, the theory is sometimes referred to as khronometric theory, and while it still violates Lorentz symmetry (being only invariant under FDiffs in the unitary gauge), 
it is quadratic in both time
and space derivatives (of the spatial metric and khronon)~\cite{Blas:2010hb,Jacobson:2013xta}. We will show that  khronometric theory can be modified to include second time derivatives of the khronon (i.e.~time derivatives of the lapse), while still propagating no \Ostro ghost.
Although similar constructions were already obtained in \cite{Gao:2018znj, DeFelice:2018mkq, Langlois:2020xbc}, here we additionally show that the resulting theory is invariant 
under spatial diffeomorphisms and a special set of (monotonic) time reparametrizations (which will turn out to be given by ``hyperbolic'' time compactifications). While invariance under this special group of transformations
is sufficient to determine the form of the kinetic term for the lapse, it does not fix
its coefficient unambiguously. Therefore, the radiative stability of the fine-tuning of the coupling constants needed to eliminate the ghost remains an open issue. We will comment on promising ways forward on this issue in the following.

As a result of this construction, the novel ``degenerate Ho\v rava gravity'' theory that we find does not have a limit to General Relativity (GR). However, somewhat surprisingly, this does not prevent the theory from having the correct Newtonian limit, nor from reproducing (at least for specific values of the coupling constants) the dynamics of GR at the first post-Newtonian (1PN) order and thus passing solar system tests. However, the behavior on cosmological scales is wildly different from GR, at least at the background level (i.e.~assuming isotropy and homogeneity). We will discuss this issue, its implications, and possible solutions below.

This paper is organized as follows. In Sec.~\ref{sec:horava} we review Ho\v rava gravity  both in the unitary gauge and in the covariant (Stuckelberg) formalism.
In Sec.~\ref{sec:rdiff} we then consider theories invariant under a smaller gauge group, i.e.~restricted foliation-preserving diffeomorphisms (RFDiffs),
as an intermediate step toward Sec.~\ref{sec:hyp}, where we introduce our degenerate generalization of Ho\v rava gravity. We study its phenomenology in Sec.~\ref{sec:phenom}, and we discuss our conclusions in Sec.~\ref{sec:concl}. We utilize units in which $c=1$, except in Appendix~\ref{PN}, where we discuss the PN expansion of degenerate Ho\v rava gravity and we thus reintroduce $c$ as a book-keeping parameter.

\section{Ho\v rava gravity}\label{sec:horava}

In this section, we review the construction of Ho\v rava gravity and the subgroup of the four-dimensional diffeomorphisms under which the theory is invariant. The distinction between space and time introduces a preferred frame, and thus a preferred time coordinate, which corresponds to endowing the space-time manifold with a preferred foliation by space-like surfaces. This means that the arbitrary reparameterization of time $t \to \tilde t (t, \xx)$ is not a symmetry of the theory anymore.

The basic ingredients to describe the space-time geometry are the spatial metric $\g_{ij}$, the shift $N^i$ and the lapse function $N$ entering the $3+1$ decomposition of the four-dimensional metric~\cite{PhysRev.116.1322}
\be
\dd s^2 = - N^2 \dd t^2 + \g_{ij} (\dd x^i + N^i \dd t)(\dd x^j + N^j \dd t)\,. \label{ADMmetric}
\ee
The Ho\v rava action is built from quantities invariant under the following unbroken symmetry, which is commonly referred to as FDiffs:
\be
\xx \to \tilde \xx(t, \xx) \,, \qquad t \to \tilde t (t) \,, \label{FDiffs}
\ee
where $\tilde t (t)$ is a monotonic function of $t$. Notice that this is the largest possible unbroken gauge group that one can have, once a preferred foliation is introduced.
Under this symmetry the fields in (\ref{ADMmetric}) transform as
\be
N \to \tilde N = N \frac{\dd t}{\dd \tilde t} \,, \quad  N^i \to \tilde N^i = \left( N^j \frac{\de \tilde x^i}{\de x^j} - \frac{\de \tilde x^i}{\de t} \right) \frac{\dd t}{\dd \tilde t} \,, \quad \g_{ij} \to \tilde \g_{ij} = \g_{kl} \frac{\de x^k}{\de \tilde x^i} \frac{\de x^l}{\de \tilde x^j} \,. \label{transfrule}
\ee
Up to dimension six, the action takes the form~\cite{Horava:2009uw}
\be
\label{actionHL}
S=\frac{1}{16\pi G}\int \dd^3x\, \dd t\, N\sqrt{\g} \left[(1-\b)K_{ij}\,K^{ij} - (1+\l) K^2 + \a \, a_i\,a^i + {}^{(3)}{R} - \V\right] \,,
\ee
where $K_{ij}$ is the extrinsic curvature of the surfaces of constant time\footnote{Note that our definition of $K_{ij}$ differs 
by an overall sign from the definition used in some textbooks (e.g. \cite{alcu,baum}), although it agrees e.g. with \cite{Blas:2009qj,Blanchet:2011wv,Bonetti:2015oda}.}
\be
K_{ij} = \frac{1}{2N} \left( \dot \g_{ij} - D_i N_j - D_j N_i \right) 
\ee
(with $D_i$ three-dimensional covariant derivatives
compatible with the spatial metric $\gamma_{ij}$ and with an overdot denoting
partial time derivatives),
$K\equiv K_{ij} \g^{ij}$ is the trace of the extrinsic curvature, ${}^{(3)}{R}$ is the three-dimensional Ricci scalar and
\be
a_i\equiv N^{-1} \de_i N
\ee
is the acceleration vector.
Besides the bare gravitational constant $G$, there are three free dimensionless constants: $\a, \b$ and $\l$.
The ``potential'' $\V$ depends on the three-dimensional Ricci tensor ${}^{(3)}{R}_{ij}$ and the acceleration $a_i$, with all possible operators of dimension four and six. This potential therefore involves only a finite number of these operators, which were fully classified e.g.
in \cite{Sotiriou:2009bx} and \cite{Blas:2009qj}. While crucial for renormalizability, for our purposes the potential is completely irrelevant. Therefore, we omit to write explicitly its form here,
and we focus on the IR limit of the theory (obtained by neglecting $\V$) in the rest of this paper. 

Also notice that the kinetic part of the action (i.e.~the part where time derivatives appear) is fully contained in the first two terms of Eq.~(\ref{actionHL}). In addition to the helicity-2 modes of the graviton, there is also a propagating scalar field, usually referred to as the ``khronon'' \cite{Blas:2010hb}. This extra mode appears because the two first-class constraints (primary and secondary) associated with time diffeomorphisms in GR become here second-class constraints, because of the breaking of time diffeomorphisms. The theory therefore possesses six first-class constraints (associated with spatial diffeomorphisms) and two second-class constraints, leaving $[20-(6\cdot 2)-2]/2 = 3$ dynamical degrees of freedom.

\subsection{Stuckelberg formalism}
\label{sec:horava_stuck}

This formalism allows one to single out explicitly the extra degree of freedom that appears because of the breaking of diffeomorphism invariance. It amounts to rewriting the action in a generally covariant form, at the expense of introducing a compensator field that transforms non-homogeneously under the broken part of the four-dimensional diffeomorphisms.

In more detail, one encodes the foliation structure of the space-time in a scalar field $\j$, such that the foliation surfaces are identified with those of constant $\j$. The action (\ref{actionHL}) then corresponds to the frame where the coordinate time coincides with $\j$ (i.e.~$\j=t$). This choice of coordinates is often referred to as ``unitary gauge''. The action in a generic frame is then obtained by performing the Stuckelberg transformation and reads
\be\label{aetheraction}
S=\frac{1}{16\pi G} \int \dd^4x\, \sqrt{-g} \left[R - \b\, \De_\m u^\n \De_\n u^\m - \l\, (\De_\m u^\m)^2 + \a\, a_\m a^\m\right] \,,
\ee
where
\be \label{khaether}
u_\mu= \frac{\de_\mu \j}{\sqrt{- X}}\,, \qquad X \equiv \de_\m \j \de^\m \j \,, \qquad a^\m\equiv u^\n \De_\n u^\mu \,,
\ee
and $R$ is the four-dimensional Ricci scalar. 
Notice that this is the action of Einstein-\ae ther theory when the \ae ther vector field is hypersurface orthogonal \cite{Jacobson:2010mx}.
For later purposes, it is convenient to write the action (\ref{aetheraction}) explicitly in terms of the khronon field
\bea
S=\frac{1}{16\pi G} \int \dd^4x\, \sqrt{-g} && \left[R + \b\, \frac{\j_{\mu \nu} \j^{\mu \nu}}{X} + \l\, \frac{(\Box \j)^2}{X} - 2\l \, \frac{(\Box \j) \j^{\mu} \j_{\mu \nu} \j^{\nu}}{X^2} \right. \nb \\
&& \left. + (\a-2\b) \frac{\j^{\mu} \j_{\mu \rho} \j^{\rho \nu} \j_{\nu}}{X^2}  + (\b+\l-\a) \frac{(\j^{\mu} \j_{\mu \nu} \j^{\nu})^2}{X^3} \right] \,, \label{khrononaction}
\eea
where to avoid clutter we have introduced the notation $\j_\m \equiv \de_\m \j$ and $\j_{\m\n} \equiv  \De_\n \de_\m \j$.
Notice that the  action above is invariant under  reparameterizations of $\j$,
\be
\j \to \tilde \j = f(\j) \,, \label{CovFDiffs}
\ee
where $f$ is a (monotonic) arbitrary function.  This reflects the invariance (in the unitary gauge) under the FDiffs (\ref{FDiffs}).

Naively, the higher-order derivatives in the action (\ref{khrononaction}) would suggest the presence of an \Ostro ghost in the theory. However, the counting of the degrees of freedom
cannot be  straightforwardly performed from the covariant action (\ref{khrononaction}),
because of the absence of a standard kinetic term for the khronon and the non-local $1/X$ dependence. 
However, one  can easily perform the counting in the preferred frame, where $\j$ cannot be constant and has a non-vanishing time profile $\j=t$. In this frame, as can be seen in the unitary gauge action (\ref{actionHL}), the ghost mode is absent.

From the point of view of the covariant action (\ref{khrononaction}),
the absence of the \Ostro mode is guaranteed by the highly non-trivial tuning among the coefficients of the five operators in the action, which translates in the absence of $\dot N$ terms in the unitary gauge action (\ref{actionHL}).
Remarkably, this tuning is protected against radiative corrections by the reparameterization invariance (\ref{CovFDiffs}). A detuning of the action coefficients would necessary break the symmetry (\ref{CovFDiffs}), generate $\dot N$ terms in the unitary gauge action, and generically reintroduce the ghost mode.

Finally, notice that the action (\ref{khrononaction}) does not belong to any of the DHOST classes identified in \cite{Crisostomi:2016czh, Achour:2016rkg}. This is because when the full diffeomorphism invariance is broken, the degeneracy of the Hessian matrix of the velocities can be achieved in a less restrictive way. The full diffeomorphism invariant analog of action (\ref{khrononaction}) is the Horndeski Lagrangian \cite{Horndeski:1974wa}, which in the unitary gauge does not present $\dot N$ terms.

\section{RFDiff gravity}\label{sec:rdiff}

Noticeably, there exists also a smaller unbroken gauge group according to which we can construct our Lagrangian, namely the group of RFDiffs
\be
\xx \to \tilde \xx(t, \xx) \,, \qquad t \to \tilde t = t + \text{const} \,, \label{RFDiffs}
\ee
which differs from FDiffs because the invariance under general time reparametrizations
is replaced by the invariance under time translations.
In the Stuckelberg formulation, this symmetry of the khronon action  reduces to the shift symmetry
\be
\j \to \tilde \j = \j + \text{const} \,,
\ee
which allows for a general dependence of the action on the derivatives of $\j$.

Restricting the symmetry from FDiffs to RFDiffs therefore allows one to include in the action a kinetic term for the lapse $N$. Moreover, all dimensionless couplings in the Lagrangian may now acquire an arbitrary dependence on $N$, and we can thus include in the potential $\V$ a generic function of $N$.
The kinetic term for $N$ is fixed by the invariance under RFDiffs to be of the form
\be
\left(\dot N - N^i \de_i N\right)^2 \,. \label{Ndot}
\ee 
However, a general dependence of the action on this term  inevitably leads to the propagation of an additional ghost scalar degree of freedom \cite{Blas:2010hb}. In the Stuckelberg formulation, this is the \Ostro mode associated with the higher derivatives of the khronon, which re-appears because of the detuning of the coefficients of the action (\ref{khrononaction}).

\subsection{Degenerate RFDiff gravity and its downsides}

It is  well established that if the kinetic term (\ref{Ndot}) and the trace of the extrinsic curvature appear in the Lagrangian in such a way as  to enforce the existence of a primary constraint (which in turn generates a secondary constraint), then the ghost mode can be safely removed \cite{Langlois:2015cwa, Langlois:2015skt, Crisostomi:2016tcp, Crisostomi:2016czh}. The constraints structure -- and so the number of degrees of freedom -- becomes indeed the same as in Horava gravity, with the only difference that the second-class primary constraint given by the vanishing of the  momentum conjugate to $N$ (i.e. $\de L/\de \dot{N} \approx 0$) is replaced by a linear combination of the momenta conjugate to $N$ and $\g$.

Therefore, it is possible to realize healthy theories within RFDiff gravity, provided that suitable degeneracy conditions are imposed. 
These models have been fully classified in \cite{Gao:2018znj, DeFelice:2018mkq, Langlois:2020xbc}, but they may not be very attractive for two reasons. First, the presence of arbitrary functions of $N$ in the Lagrangian results in an infinite number of coupling constants. Second,  the degeneracy conditions are not protected by the RFDiff symmetry, so that radiative corrections will generically induce a detuning of the action and hence reintroduce the ghost. This is very different from Ho\v rava gravity, where  the tuning of the action \eqref{khrononaction} is required by the FDiff symmetry and hence is protected by it.

\section{Degenerate Ho\v rava gravity}\label{sec:hyp}

In this section, we present a new class of gravity theories 
invariant under a symmetry intermediate
 between  FDiffs and  RFDiffs. 
In more detail, the transformation of time is restricted to take the form of a specific hyperbolic function $\tilde t (t)$, and the symmetry is realized up to a total derivative.

Our starting point is a generalization of the Ho\v rava action (\ref{actionHL}), which includes the time derivative of the lapse in the RFDiff invariant way (\ref{Ndot}). By introducing the definition
\be
V \equiv - \frac{1}{N^2} \left( \dot N - N^i \de_i N \right) \,,
\ee
we write the action as
\be
\label{actionNEW}
S=\frac{1}{16 \pi G}\int \dd^3x\, \dd t\, N\sqrt{\g} \left[  \o \, V^2 + 2\, \s \, K \, V + (1-\b)K_{ij}\,K^{ij} - (1+\l) K^2 + \a \, a_i\,a^i + {}^{(3)}{R} \right] \,,
\ee
where $\o$ and $\s$ are two additional dimensionless constants. 
Clearly, the first two terms in (\ref{actionNEW}) break FDiff invariance, although they are RFDiff invariant.
At this stage, two scalar degrees of freedom propagate, one of which is a ghost \cite{Blas:2010hb}.

We can then impose the existence of a primary constraint by requiring that the determinant of the kinetic matrix of the two scalar modes in (\ref{actionNEW}) vanishes \cite{Motohashi:2016ftl, Klein:2016aiq, Crisostomi:2017aim}:
\be
\text{det} \left(
\begin{matrix} 
    \o \qquad & \s \\
    \s \qquad & - \l - \frac{\b+2}{3}
\end{matrix}
\right) = 0 \,. \label{condM}
\ee
In Ho\v rava gravity, this condition is trivially realized since $\o=\s=0$,
but a non-trivial solution is also possible and is given by
\be
\o = - \frac{3\,\s^2}{3\,\l +\b +2} \,. \label{cond}
\ee
To completely remove one degree of freedom, the primary constraint, enforced by the condition (\ref{cond}), must generate a secondary constraint. The conditions for this to happen were derived in complete generality for field theories in \cite{Crisostomi:2017aim}, and in the case at hand they are automatically satisfied because of the absence of the following couplings in the action:
\be
V \de_i N \,,\qquad  K \de_i N \,, \qquad {}^{(3)}{R} \cdot V \,, \qquad {}^{(3)}{R} \cdot K \,.
\ee
Therefore, condition (\ref{cond}) is all that is needed to completely eliminate the ghost mode and remain with a single scalar field, the khronon.

Thus far, we have not made any progress with respect to degenerate RFDiff theories, of which the action (\ref{actionNEW}) -- with the condition (\ref{cond}) enforced -- is a particular case. In fact, nothing prevents the couplings in the Lagrangian from being functions of $N$ and, more dangerously, quantum corrections from spoiling the condition (\ref{cond}).
Ideally, our aim is to determine whether there exists a gauge group, smaller than the FDiffs one but larger than the RFDiffs one, that could protect the condition (\ref{cond}). 

For this purpose, after imposing the condition (\ref{cond}), we transform the action (\ref{actionNEW}) under the FDiffs  (\ref{FDiffs}) and obtain a new action,
which  now differs from the original one, because $V$ is not invariant. Indeed, using Eq.~(\ref{transfrule}), we find that
\be
V \to \tilde V = V - \frac{1}{N} \frac{\dd^2 t}{\dd \tilde t ^2} \left(\frac{\dd t}{\dd \tilde t}\right)^{-2} \,. \label{Vtransf}
\ee
By requiring that the new terms generated by Eq.~(\ref{Vtransf}) in the action are a total derivative, one
imposes that the equations of motion are invariant. In this way, we obtain a third-order differential equation for $\tilde t (t)$, which has a non-trivial solution only if the following condition is imposed on the coefficients of the Lagrangian
\be
\s = - \l - \frac{\b + 2}{3} \,. \label{cond2}
\ee
In this case, there exists a unique family of solutions for $\tilde t (t)$ given by
\be
\tilde t (t) = \frac{c_2}{c_1 + t} + c_3 \,, \label{HFDiffs}
\ee
where $c_{1,2,3}$ are free integration constants. Being (\ref{HFDiffs}) a hyperbolic function, we refer to this family of time reparametrizations (together with spatial diffeomorphisms) as  ``hyperbolic-foliation-preserving Diffs'' (HFDiffs). Notice that Eq.~\eqref{HFDiffs} is a monotonic function on
intervals not including its pole, which prevents violations of causality.

By inspection of the resulting Lagrangian, it is easy to realize that the conditions (\ref{cond}) and (\ref{cond2}) force the kinetic term to be 
\be
\left( V + K \right)^2 \,, \label{kinform}
\ee
which  is indeed invariant under HFDiffs up to a total derivative since
\be
N\sqrt{\g} \left(  V + K \right)^2 \to N\sqrt{\g} \left(  V + K \right)^2 - 4\left[ \de_t \left( \frac{\sqrt{\g}}{N(c_1 +t)} \right) - \de_i \left( \frac{\sqrt{\g} N^i}{(c_1+t)N} \right) \right] \,.
\ee
However, being $N\sqrt{\g} K^2$ FDiff invariant, if one starts from the generic
Lagrangian \eqref{actionNEW}, invariance under HFDiffs (up to a total derivative)
only requires the kinetic term to be of the form $V^2+2 K V$, with an arbitrary coefficient in front. Therefore, this implies
that radiative corrections may induce a running of the coefficients of the Lagrangian.  This can potentially detune the degeneracy condition 
\eqref{cond} and thus reintroduce the ghost.

A way out of this unappealing situation would be to
enlarge the group of HFDiffs to a
custodial symmetry sufficient  to
ensure stability of the degeneracy condition under quantum corrections. Such a
gauge group must necessarily lie in between
HFDiffs and full-fledged four-dimensional diffeomorphisms (since
the latter would simply require the theory to be GR). Therefore,
one might consider a specific class of time reparametrizations that are explicit 
functions of the spatial coordinate, e.g. $\tilde t (t, \xx) = {c_2}/(c_1 + t) + c_3 + f(\xx)$, although we have been unable to identify a suitable function $f(\xx)$
of the spatial coordinates.

To summarize, we have found a new unbroken gauge group that: \textit{(i)} allows for a kinetic term for the lapse; \textit{(ii)}  avoids
the presence of arbitrary functions of the lapse in the action; although \textit{(iii)} it
does not yet prevent the propagation of a ghost mode.
These are the HFDiffs
\be
\xx \to \tilde \xx(t, \xx) \,, \qquad t \to \tilde t = \frac{c_2}{c_1 + t} + c_3 \,, \label{HFDiffs2}
\ee
and the corresponding action is given by
\bea
S=\frac{1}{16 \pi G}\int \dd^3x\, \dd t\, N\sqrt{\g} && \left[ - \left( \l + \frac{\b + 2}{3} \right) \left( V^2 + 2 \, K\, V \right) \right. \nb\\
&& \left. + (1-\b)K_{ij}\,K^{ij} - (1+\l) K^2 + \a \, a_i\,a^i + {}^{(3)}{R} \right] \,. \label{actionNEW2}
\eea
Comparing with the Ho\v rava action (\ref{actionHL}) we notice that, although there are two new operators, 
the number of coupling constants is the same. In the this new action, however, even when  all couplings are set to zero, we do not recover the GR limit. 

\subsection{Stuckelberg formalism}

It is now instructive to look at the new action (\ref{actionNEW2}) in the Stuckelberg formalism.
As in Section \ref{sec:horava_stuck}, we perform the Stuckelberg transformation and obtain
\bea
S=\frac{1}{16 \pi G} \int \dd^4x\, \sqrt{-g} && \left[R + \b\, \frac{\j_{\mu \nu} \j^{\mu \nu}}{X} + \l\, \frac{(\Box \j)^2}{X} + \frac{2(\b+2)}{3} \, \frac{(\Box \j) \j^{\mu} \j_{\mu \nu} \j^{\nu}}{X^2} \right. \nb \\
&& \left. + (\a-2\b) \frac{\j^{\mu} \j_{\mu \rho} \j^{\rho \nu} \j_{\nu}}{X^2}  + \frac{(2\b - 3\a - 2)}{3} \frac{(\j^{\mu} \j_{\mu \nu} \j^{\nu})^2}{X^3} \right] \,. \label{khrononnewaction}
\eea
Comparing with the khronon action for Ho\v rava gravity, Eq.~(\ref{khrononaction}), we see that the coefficients of the third and fifth operators have changed. This new highly non-trivial tuning guarantees the absence of the \Ostro ghost at least at tree level.
Moreover, the Lagrangian \eqref{khrononnewaction}
is invariant under the transformation
\be
\j \to \tilde \j = \frac{c_2}{c_1 + \j} + c_3 \,, \label{CovHFDiffs}
\ee
up to the total derivative
\be
- 4 \left( \l + \frac{\b + 2}{3}  \right) \De_\m \left(\frac{1}{c_1+\j} \de^\m \j \right) \,.
\ee
This reflects the invariance, up to a total derivative and in the unitary gauge, under the HFDiffs (\ref{HFDiffs2}).

Finally, notice that also in this case, action (\ref{khrononnewaction}) does not belong to any of the DHOST classes \cite{Crisostomi:2016czh, Achour:2016rkg}. Although the end result is the same as in the DHOST and beyond Horndeski constructions (i.e. higher order equations of motion without \Ostro ghost),
there is no connection between those theories and ours. 
In fact,  our theory is Lorentz-violating and the absence of ghosts is guaranteed precisely by the existence of a preferred foliation, 
where kinetic terms take their standard form and 
the  non-local $1/X$ terms present in the Stuckelberg action disappear.


\subsection{Conformal and disformal transformation}

Looking at the form of the kinetic terms, Eq.~(\ref{kinform}), a natural question  is whether the new theory is related to Ho\v rava gravity by a conformal and/or disformal transformation.
To check this, it is convenient to work in the Stuckelberg formalism, where these transformations read \cite{Bekenstein:1992pj}
\be
\bar{g}_{\mu \nu} = \Omega(X) g_{\mu \nu} + \Gamma(X) \de_\m \j \, \de_\n \j \,, \label{transf}
\ee
with $\Omega$ and $\Gamma$  free functions of $X$ only. (Notice that a $\j$ dependence would break even the shift symmetry and therefore it is not allowed).

It is well known that Ho\v rava gravity is invariant under the transformation given by \cite{Foster:2005ec} 
\be
\Omega =1 \,, \qquad \Gamma = \frac{\varsigma}{X} \,, \label{confdisf}
\ee 
where $\varsigma$ is a constant, provided the following rescaling of the coefficients
\be
\bar \l = \l + \varsigma \left(\l +1\right) \,, \qquad \bar \b = \b +\varsigma \left(\b-1\right) \,. \label{rescaling}
\ee
A  feature of the new theory 
is that it also enjoys this invariance for the very same choice of functions [Eq.~(\ref{confdisf})] and for the same rescaling of the coefficients [Eq. (\ref{rescaling})]. Moreover, any choice different from Eq.~(\ref{confdisf}) would change the power of $X$ appearing in each of the operators of the Lagrangian, and therefore cannot connect the two theories.

As a consequence, a generic transformation of the form (\ref{transf}) cannot map Ho\v rava gravity into its degenerate HFDiff generalization, and both theories are stable under the same transformation (\ref{confdisf}).
Once again, the non-local form of $\Gamma$ in (\ref{confdisf}) is signaling that the two theories make sense only for $X \neq 0$, i.e.~the khronon must always be timelike.

\section{Phenomenology}\label{sec:phenom}

To study the phenomenology of ``degenerate'' HFDiff  Ho\v rava gravity, we  first derive the field equations by varying the action \eqref{actionNEW2} with respect to the spatial metric $\gamma_{ij}$, the shift $N_i$ and the lapse $N$.
We denote by $D_i$ the covariant derivative defined with respect to $\gamma_{ij}$, and $D_t\equiv\partial_t - N_k D^k$. We also define the following quantities in terms of the variation of the matter action~\cite{Blanchet:2011wv,Bonetti:2015oda}
\begin{align}
\mathcal{E}&=-\frac{1}{\sqrt{\gamma}}\frac{\delta S_{\mathrm{m}}}{\delta N}=N^2 T^{00}\;,\\
\mathcal{J}^i&=\frac{1}{\sqrt{\gamma}}\frac{\delta S_{\mathrm{m}}}{\delta N_i}=N\left(T^{0i}+N^iT^{00}\right)\;,\\
\mathcal{T}^{ij}&=\frac{2}{N\sqrt{\gamma}}\frac{\delta S_{\mathrm{m}}}{\delta \gamma_{ij}}=T^{ij}-N^iN^jT^{00}\;,
\end{align}
where $T^{\mu\nu}$ is the four-dimensional matter energy-momentum tensor.

The variation with respect to $N$ leads to 
\begin{align}
\frac{^{(3)}R}{1-\beta}+\frac{\lambda+1}{1-\beta}&K^2 -K_{ij}K^{ij}+\frac{\alpha (D_i N)(D^i N)}{(1-\beta)N^2}-\frac{2\alpha D^i D_i N}{(1-\beta)N}\nonumber \\&+\frac{2(2+\beta + 3\lambda)}{3(1-\beta)}\left[KV+V^2-K^2-D_t\left(\frac{K+V}{N}\right)\right]=\frac{16 \pi G\mathcal{E}}{(1-\beta)c^4}\;,\label{eq:H}
\end{align}
which unlike in GR is not a constraint, but rather an evolution equation for $N$
(c.f.~the presence of both second-order space and time derivatives of $N$). Notice that in (non-degenerate) Ho\v rava gravity\footnote{We often refer to the original Ho\v rava gravity [Eq.~(\ref{actionHL})] as ``non-degenerate'', in order to distinguish it from its degenerate extension [Eq.~(\ref{actionNEW2})]. However, it should be by now well understood that even non-degenerate Ho\v rava gravity is a degenerate theory that satisfies (although trivially) the degeneracy condition (\ref{condM}). This is even more evident from the tuning of the parameters in its khronometric formulation~(\ref{khrononaction}).},
this equation is instead an elliptic equation for $N$, to be solved on each slice~\cite{Blanchet:2011wv,Bonetti:2015oda}. In the covariant formalism, this equation becomes indeed (in both non-degenerate and degenerate Ho\v rava gravity) the khronon evolution equation.
Varying with respect to $N_i$ one obtains the momentum constraint equation
\begin{align}
D_j\Bigg[\bigg(K^{ij}-\frac{\lambda+1}{1-\beta}&\gamma^{ij}K\bigg)-\frac{(2+\beta + 3\lambda)\gamma^{ij}V}{3(1-\beta)}\Bigg]\nonumber \\&-\frac{(2+\beta + 3\lambda)}{3(1-\beta)}\left(D^i N\right)\left(\frac{K+V}{N}\right)=-\frac{8\pi G\mathcal{J}^i}{(1-\beta)c^4}\;,\label{eq:Hi}
\end{align}
while variation with respect to $\gamma_{ij}$ yields the evolution equation
\begin{align}
&\frac{1}{1-\beta}\left[\;^{(3)}R^{ij}-\frac{1}{2}\;^{(3)}R \gamma^{ij}\right]+\frac{1}{N}D_t \left(K^{ij}-\frac{\lambda+1}{1-\beta} K \gamma^{ij}-\frac{(2+\beta+3\lambda)}{3(1-\beta)}V\gamma^{ij}\right)\nonumber \\ &+\frac{2}{N}D_k\left(N^{(i}[K^{j)k}-\frac{\lambda+1}{1-\beta}K\gamma^{j)k}-\frac{(2+\beta+3\lambda)}{3(1-\beta)} V \gamma^{j)k}]\right)+2K^{ik}K_{k}^j-\frac{2\lambda+1+\beta}{1-\beta}K K^{ij}\nonumber \\ & - \frac{1}{2}\gamma^{ij}\left(K_{kl}K^{kl}+\frac{\lambda+1}{1-\beta}K^2\right)-\frac{1}{(1-\beta)N}\left[(D^iD^j N)-(D_k D^k N)\gamma^{ij}\right]\nonumber \\ &+\frac{\alpha}{N^2(1-\beta)}\left[(D^i N)(D^j N)-\frac{1}{2}(D_k N)(D^k N)\gamma^{ij}\right]\nonumber \\&+\frac{2(2+\beta+3\lambda)}{3(1-\beta)}\left[\frac{1}{4} V^2\gamma^{ij}-K^{ij}V -  \frac{K+V}{N^2}N^{(i} (D^{j)} N)\right]=\frac{8 \pi G}{(1-\beta)c^4}\mathcal{T}^{ij}\;.\label{eq:ev}
\end{align}

\subsection{Solar-system tests and gravitational-wave propagation}
To check the experimental viability of the theory on Earth and in the solar system, we perform a post-Newtonian expansion over flat space and compare to the parametrized PN metric (PPN)~\cite{Will:1993hxu,Will:2014kxa}.
This will allow us to extract the values of the PPN parameters in degenerate 
Ho\v rava gravity and compare them to their experimental bounds. 
The details of the calculation follow \cite{Bonetti:2015oda}, which performs the same analysis for (non-degenerate) Ho\v rava gravity, and are presented 
in Appendix~\ref{PN}. Here, we will simply summarize the main results.

We find that the only  PPN parameters differing from GR are the preferred-frame parameters $\alpha_1$ and $\alpha_2$, which take the form
\begin{align}
\alpha_1&=\frac{4(\alpha-2\beta)}{\beta-1}\;,\\
\alpha_2&=-\frac{(\alpha -4 \beta +2) (3 \alpha -4 \beta -2)}{3(\alpha-2)(\beta +\lambda) }+\frac{2(\alpha-2)}{\beta-1}+\frac{-27 \alpha +28 \beta +12
   \lambda +38}{3 (\alpha -2)}\;.\label{alpha2}
\end{align}
Experimental bounds on these parameters are
$|\alpha_1|\lesssim 10^{-4}$ and $|\alpha_2|\lesssim 10^{-7}$~\cite{Will:2014kxa}.
Comparing to their expressions in (non-degenerate) Ho\v rava gravity \cite{Blas:2011zd,Bonetti:2015oda}, we see that while $\alpha_1$ is unchanged,  $\alpha_2$ gets modified. In (non-degenerate) Ho\v rava gravity, $\alpha_1$ and $\alpha_2$ are both proportional to $\alpha-2\beta$, i.e.~they are both small for $\alpha\approx 2\beta$. 

At this point, let us notice that constraints on the propagation speed of gravitational waves from GW170817 require $|\beta|\lesssim 10^{-15}$ in (non-degenerate) Ho\v rava gravity~\cite{Monitor:2017mdv,Gumrukcuoglu:2017ijh}, as well as in the degenerate version of the theory that we are considering here.
Indeed, the kinetic term of the tensor modes is given by $K_{ij}\,K^{ij}$ and the spatial gradient is contained in ${}^{(3)}{R}$, 
which gives [c.f. Eqs.~(\ref{actionHL}) and~(\ref{actionNEW2})] a gravitational-wave propagation speed $c_{\rm GW}=(1-\beta)^{-1/2}$, which matches the speed of light only for $\beta=0$.
 
The solar-system bound on $\alpha_1$ then gives $\alpha\approx \beta\approx 0$ in both non-degenerate~\cite{ramos,Barausse:2019yuk}
and degenerate Ho\v rava gravity.
For $\alpha\approx \beta\approx 0$, Eq.~\eqref{alpha2} then yields
\begin{equation}
\alpha_2\approx-\frac{(1+2\lambda)(2+3\lambda)}{3\lambda}\;.
\end{equation}
 The experimental constraint
  $|\alpha_2|\lesssim 10^{-7}$ then selects $\lambda\approx-1/2$ or $\lambda\approx-2/3$.
  For the latter, the coefficient in front of $V^2+2KV$ in the action disappears, i.e.~one is left with the 
non-degenerate version of the theory. Therefore, there exists only one non-trivial
set of parameters, namely $\alpha\approx \beta \approx0$ and $\lambda\approx-1/2$, for which degenerate Ho\v rava gravity can satisfy solar-system tests and the bound
on the propagation speed of gravitational waves. For these values the action  reads
\bea
S=\frac{1}{16\pi G} \int \dd^4x\, \sqrt{-g} \left[R - \frac{1}{2} \frac{(\Box \j)^2}{X} + \frac{4}{3} \, \frac{(\Box \j) \j^{\mu} \j_{\mu \nu} \j^{\nu}}{X^2} 
- \frac{2}{3} \frac{(\j^{\mu} \j_{\mu \nu} \j^{\nu})^2}{X^3} \right] \, \label{khrononphenoaction}
\eea
or, in the unitary gauge,
\begin{align}
S=\frac{1}{16\pi G}\int \ud^3x\, \ud t\, N\sqrt{\gamma} & \left[ -\frac16 \left( V^2 + 2 \, K\, V \right)  + K_{ij}\,K^{ij} - \frac12 K^2   + {}^{(3)}{R} \right] \,.
\end{align}

\subsection{Cosmology}

To test the behavior of the theory on cosmological scales, we assume a standard homogeneous and isotropic Robertson-Walker metric
\begin{equation}
\ud s^2 = -\ud t^2 + a^2(t)\delta_{ij}\ud x^i \ud x^j\;,
\end{equation}
where $a(t)$ is the scale factor, and $\gamma_{ij}=a^2(t) \delta_{ij}$.
Notice that we have assumed  flat spatial slices, but our conclusions are unchanged if we allow for curvature. 

Replacing this ansatz in the field equations, the
only non-trivial equations are provided by the khronon equation \eqref{eq:H}
and by the trace of the evolution equation \eqref{eq:ev}, which give
the system
\begin{align}\label{frw1}
(2+\beta+3\lambda)(3H^2+2\dot{H})&=-16\pi G \rho\;,\\
(2+\beta+3\lambda)(3H^2+2\dot{H})&=-16\pi G P\;,\label{frw2}
\end{align}
with $H=\dot{a}/a$ the Hubble rate, while $\rho$ and $P$ are the energy density and pressure of the cosmic matter.

As can be seen, this system is completely different from the Friedmann-Robertson-Walker equations of GR. This is no surprise since
the theory does not reduce to GR for any values of the coupling constants. More worrisome is the fact that by taking the difference of the two equations, one obtains that the cosmic matter  must necessarily have $\rho=P$
(stiff fluid). In other words, the theory does not allow for the usual radiation and matter eras, nor for an early- or late-time accelerated expansion (even in the presence of a cosmological constant). As a curiosity, however, it is worth mentioning that if we set $\rho=P$ in Eqs.~\eqref{frw1}--\eqref{frw2} and solve for $H$, we find $H(t)=2/[3 (t+C)]$, with $C$ an integration constant. For $C=0$ this reduces to the Hubble rate of the standard matter-dominated era, and is reminescent of the appearance of Dark Matter as an integration constant in projectable Ho\v rava gravity~\cite{Mukohyama:2009mz}.

\section{Conclusions}\label{sec:concl}

In this work, we have shown that it is possible to construct a novel khronometric theory with a dynamical lapse, which (via a degenerate Lagrangian) propagates only a graviton and a khronon. This theory is invariant under a special subgroup of the FDiff symmetry, Eq.~\eqref{HFDiffs2}, which we have referred to as hyperbolic-foliation-preserving Diffs (HFDiffs).
This new unbroken gauge group selects a specific kinetic term for the lapse (although it does not fix its overall coefficient), and it avoids an arbitrary dependence of the action on the lapse. 
HFDiffs are not sufficient by themselves to
ensure stability of the degeneracy condition under radiative corrections, thus
potentially letting the ghost re-appear  beyond tree level. However, an enlarged
gauge group lying between HFDiffs and four-dimensional diffeomorphisms 
may protect the fine-tuning of the degeneracy condition. We have commented on
this possibility above, and we will explore it further in future work.

Our construction has a two-fold interest for both phenomenology and theory. On the phenomenological side, it is a remarkable example of a theory which, despite being Lorentz breaking and not admitting a GR limit,  does pass
 Earth-based, solar-system and gravitational-wave tests, at least for a suitable choice of its coupling constants. 
Unfortunately, the theory fails to reproduce the standard Friedmann-Robertson-Walker cosmology and can provide an (effective) matter-dominated era only if the universe contains stiff matter alone ($\rho=P$). 
However, while clearly the cosmology of the theory does not seem to work out of the box, a couple possibilities are worth mentioning. First,  Eqs.~\eqref{frw1}--\eqref{frw2} assume a minimal coupling to matter. If matter is instead conformally coupled to gravity, it may be possible to obtain a matter-dominated era and a late-time accelerated expansion, although it would still be impossible to accommodate a radiation era and it might be tricky to pass solar-system tests (at least in the absence of a screening mechanism protecting local scales from the conformal coupling). Second, and perhaps more importantly, since dark matter seems to arise naturally as an integration constant in our
new theory, it may be worth trying to explain the observed late-time acceleration of the universe in the context of non-standard cosmologies that violate the homogeneity/isotropy assumptions of the Robertson-Walker ansatz (see e.g. \cite{Clarkson:2011zq} for a review, and references therein).

On the theoretical side, if a custodial symmetry
protecting the degeneracy condition is identified, this theory may
provide a version of Ho\v rava gravity that does not require the absence of time derivatives of the lapse to avoid ghosts, and hence may not present the same technical hurdles \cite{Barvinsky:2015kil,Barvinsky:2017zlx} in proving perturbative renormalizability (beyond power counting) that one encounters
in FDiff (i.e.~non-degenerate) Ho\v rava gravity 
(at least in its general non-projectable form).

\acknowledgments

We thank M. Herrero-Valea for insightful conversations on Ho\v rava gravity and its perturbative renormalizability. We also thank S. Sibiryakov for pointing out an error in the first version of this manuscript.
E.B, M.C. and L.t.H acknowledge financial support provided under the European Union's H2020 ERC Consolidator Grant ``GRavity from Astrophysical to Microscopic Scales'' grant agreement no. GRAMS-815673. S.L. acknowledge funding from the Italian Ministry of Education and Scientific Research (MIUR) under the grant PRIN MIUR 2017-MB8AEZ.

\appendix

\section{Post-Newtonian Expansion}\label{PN}
In order to calculate the PPN parameters, we follow \cite{Bonetti:2015oda} and consider a general perturbed flat metric in Cartesian coordinates $(x^0=ct,x^i)$
\begin{align}\label{eq:perturbed flat metric}
g_{00} &= -1 -\frac{2}{c^2}\phi - \frac{2}{c^4}\phi_{\mbox{\tiny{(2)}}} + {\cal O} \left(\frac{1}{c^6}\right)\notag\;,\\
g_{0i} &= \dfrac{w_i}{c^3} + \dfrac{\partial_i \omega}{c^3} + {\cal O} \left(\frac{1}{c^5}\right)\notag\;,\\
g_{ij} &= \left(1 - \frac{2}{c^2}\psi\right)\delta_{ij} + \left(\partial_i \partial_j - \frac{1}{3}\delta_{ij}\nabla^2\right)\dfrac{\zeta}{c^2}\notag\\ 
&+ \dfrac{1}{c^2}\partial_{(i}\zeta_{j)} + \dfrac{\zeta_{ij}}{c^2} + {\cal O} \left(\frac{1}{c^4}\right)\,,
\end{align}
where under transformations of the spatial coordinates, $\psi, \zeta, \omega, \phi, \phi_{\mbox{\tiny{(2)}}}$ transform as scalars, $w_i, \zeta_{i}$ behave as transverse vectors (i.e.~$\partial_i w^i = \partial_i\zeta^{i} = 0$),
and $\zeta_{ij}$ is a transverse and traceless tensor (i.e.~$\partial_i\zeta^{ij}=\zeta_{\ i}^i=0$). 
Since we want to use this ansatz in the field equations in the unitary gauge (Eqs.~\eqref{eq:H}--\eqref{eq:ev}), we are not allowed to perform a  transformation of the time coordinate (which is fixed to coincide with the khronon), but we can perform a gauge transformation of the spatial coordinates to set $\zeta=\zeta_i=0$~\cite{Bonetti:2015oda}.

We supplement this ansatz with an expression for the energy-momentum tensor, which we assume to be given by the perfect fluid form
\begin{equation}
T^{\mu\nu} = \biggl(\rho + \dfrac{P}{c^2}\biggr)u^{\mu}u^{\nu} + Pg^{\mu\nu},
\end{equation}
where $\rho$ is the matter energy density, $P$ the pressure and $u^{\mu} = dx^{\mu}/d\tau$ the four-velocity of the fluid elements (with $\tau$ the proper time). In the following, we introduce a parameter $\eta_{0,1}$ in the action
\begin{align}
S=\frac{1}{16\pi G}\int \ud^3x\, \ud t\, N\sqrt{\gamma} & \left[ -\eta_{0,1} \left( \lambda + \frac{\beta + 2}{3} \right) \left( V^2 + 2 \, K\, V \right) \right. \nb \\
& \left. + (1-\beta)K_{ij}\,K^{ij} - (1+\lambda) K^2 + \alpha \, a_i\,a^i + {}^{(3)}{R} \right] \,,
\end{align}
in order to distinguish between (non-degenerate)  Ho\v rava gravity (which 
corresponds to $\eta_{0,1}=0$) from its degenerate generalization ($\eta_{0,1}=1$).

 Expanding the evolution equation \eqref{eq:ev} to lowest order in $1/c$, we find
 $\zeta_{ij}={\cal O} (1/c^2)$ from the off-diagonal part, while the trace gives
 \begin{equation}\label{eq:quasi_equality}
\psi = \phi + {\cal O} \left(\frac{1}{c^2}\right)\,.
\end{equation}
 From this we can write
\begin{equation}
\psi = \phi +\frac{\delta\psi}{c^2}+ {\cal O} \left(\frac{1}{c^4}\right)\,,
\end{equation}
which we can substitute in the other equations. 
Using this expression and expanding Eq.~\eqref{eq:H} to lowest order in $1/c$, we obtain the modified Poisson equation
\begin{equation}
\nabla^2 \phi_N = 4 \pi G_N \rho + {\cal O} \left(\frac{1}{c^2}\right)\;,
\end{equation}
where we define the rescaled gravitational constant 
$G_N=2G/(2-\alpha)$.
Notice that $G_N$ is the gravitational constant measurable by a local experiment,
while $G$ is merely the bare gravitational constant appearing in the action.

We can then expand the momentum constraint \eqref{eq:Hi} to lowest order in $1/c$ to find the 1PN equation for the ``frame-dragging'' potential $w_i$:
\begin{align}\label{momLO}
\nabla^2 w_i + 2\left(\dfrac{\beta+\lambda}{\beta -1}\right)&\partial_i \nabla^2 \omega = \dfrac{16\pi G \rho v_i }{1-\beta} -\frac{2(3+\eta_{0,1})}{3}\left(\dfrac{2+\beta+ 3\lambda}{\beta -1}\right)\partial_i \partial_t\phi\,.
\end{align}
This is the first place where one can see a modification
with respect to (non-degenerate) Ho\v rava gravity. 

One can then expand both the trace of the evolution equation~\eqref{eq:ev}
and the khronon equation \eqref{eq:H} to next-to-leading order in $1/c$, obtaining
 respectively
\begin{align}\label{eq:delta_psi}
2\nabla^2 \delta\psi =-24\pi G p -& 8\pi\rho v^2-\left(7+\frac{\alpha}{2}\right)\partial_i \phi \partial_i \phi \nonumber\\
&- 8\phi\nabla^2\phi + \left(2 + \beta + 3\lambda\right)\left(\partial_t\nabla^2\omega + (3+\eta_{0,1})\partial_t^2 \phi\right)\,,
\end{align}
and
\begin{align}\label{eq:modified_poisson_1PN}
\vec{\nabla} \cdot \Biggl[\biggl(1-\dfrac{\alpha}{2}\biggr)\vec{\nabla} \biggl(\phi +& \dfrac{\phi_{\mbox{\tiny{(2)}}}}{c^2}\biggr)\Biggr] = 4\pi G \rho + \dfrac{1}{c^2}\Bigl(8\pi G \rho v^2 + 12\pi G p + (2-\alpha)\vec{\nabla}\phi\cdot\vec{\nabla} \phi \nonumber\\
& - \dfrac{1}{6}(2+\beta+3\lambda)\left((3+\eta_{0,1})\partial_t\nabla^2 \omega + (9+7\eta_{0,1})\partial_t^2\phi\right)\Bigr).
\end{align}

\noindent
Then, we use the same methods and notation  described in Appendix A of \cite{Bonetti:2015oda} and define the  potentials
\begin{gather}
\mathbb{X}(\vec{x}, t) =  G_N \int d^3 x' \rho(\vec{x} ', t) \ |\vec{x} - \vec{x}'|, \\ V_{i}=G_N\int d^{3}x' \ \frac{\rho(\vec{x} ', t) v'_{i}}{|\vec{x}-\vec{x} '|},
\\
 W_{i}=G_N\int d^{3}x' \ \frac{\rho(\vec{x} ', t) \ \vec{v} ' \cdot (\vec{x}-\vec{x}\ ') \ (x-x')_{i}}{|\vec{x}-\vec{x} '|}, \\ 
\Phi_{1}=G_N\int d^{3}x' \frac{\rho(\vec{x} ', t) v'^{2}}{|\vec{x}-\vec{x}'|}, \\\Phi_{2}=-G_N\int d^{3}x' \frac{\rho(\vec{x} ', t) \phi_N(\vec{x} ', t)}{|\vec{x}-\vec{x} '|}, \\ \Phi_{4}=G_N\int d^{3}x' \frac{P(\vec{x} ', t)}{|\vec{x}-\vec{x} '|},
\end{gather}
which obey the following relations
\begin{gather}\label{eq:relazioni1}
 \nabla^2 \mathbb{X} = -2\phi_N, \\ \nabla^2 V_i = -4\pi G_N \rho v_i, \\ \nabla^2\Phi_{1} = -4\pi G_N \rho v^2, \\ \nabla^2\Phi_{2} = 4\pi G_N \rho \phi_N,\\
\nabla^2\Phi_{4} = -4\pi G_N P, \\ \partial_i V^i = \partial_t \phi_N, \\ \partial_i V^i = -\partial_i W^i, \\  \partial_t\partial_i \mathbb{X} = W_i - V_i\,.\label{eq:relazioni_last}
\end{gather}

\noindent
We can then take the divergence of Eq.~(\ref{momLO}) and solve it for $\omega$, obtaining
\begin{equation}
\omega=\frac{3\alpha+2\eta_{0,1} + (\beta+3\lambda)(3+\eta_{0,1})}{6(\beta+\lambda)}\partial_t \mathbb{X}\;.
\end{equation}
Replacing this solution again into Eq.~(\ref{momLO}), we obtain
\begin{equation}
w_i=\frac{2-\alpha}{\beta-1}(V_i+W_i)\;.
\end{equation}
This allows us to evaluate $g_{0i}$ as
\begin{align}
g_{0i} &= \dfrac{w_i}{c^3} + \dfrac{\partial_i \omega}{c^3} +{\cal O}\left(\frac{1}{c^5}\right)\notag\\&= \dfrac{-\eta_{0,1}(2+3\lambda-\beta)+\beta(\beta+3\lambda)(3+\eta_{0,1})-3\alpha(1+\beta+2\lambda)+3\lambda+9\beta}{6(\beta-1)(\beta+\lambda)}\dfrac{W_i}{c^3}\notag\\
& + \dfrac{\eta_{0,1}(2+3\lambda-\beta)-\beta(\beta+3\lambda)(3+\eta_{0,1})+3\alpha(1-3\beta-2\lambda)+21\lambda+15\beta}{6(\beta-1)(\beta+\lambda)}\dfrac{V_i}{c^3} +{\cal O}\left(\frac{1}{c^5}\right)\,.
\end{align}
We can also solve Eq.~\eqref{eq:modified_poisson_1PN} for $\phi_{\mbox{\tiny{(2)}}}$
\begin{align}
\phi_{\mbox{\tiny{(2)}}} = \phi_N^2 - 2\Phi_1 - 2\Phi_2 - 3\Phi_4+ \dfrac{(3\alpha-6\beta+\eta_{0,1}(\alpha-6\beta-4\lambda))(2+\beta+3\lambda)}{6(\alpha-2)(\beta+\lambda)}\partial^2_t \mathbb{X}\,.
\end{align}

While the solutions that we found completely describe the metric at 1PN order,
in order to read off the PPN parameters one needs to transform the metric from the unitary gauge that we used for the calculation to the standard PN gauge~\cite{Will:1993hxu,Will:2014kxa,Bonetti:2015oda}. We do that, following again
\cite{Bonetti:2015oda}, by performing
 a gauge transformation $t\to t+\delta t$, where we choose
$\delta t\propto \partial_t\mathbb{X}$, with $\eta_{0,1}$ appearing in the transformation. This finally yields
\begin{align}
&g_{00} = -1 -2\dfrac{\phi_N}{c^2} -2\dfrac{\phi_N^2}{c^4} + 4\dfrac{\Phi_1}{c^4} + 4\dfrac{\Phi_2}{c^4} + 6\dfrac{\Phi_4}{c^4}+{\cal O}\left(\frac{1}{c^6}\right)\;,\\
& g_{0i} =-\dfrac{1}{2}\Bigl(7 +\alpha_1-\alpha_2\Bigr)\dfrac{V_i}{c^3} - \dfrac{1}{2}\Bigl(1 +\alpha_2\Bigr)\dfrac{W_i}{c^3}+{\cal O}\left(\frac{1}{c^5}\right)\;,\\
& g_{ij} = \left(1-2\dfrac{\phi_N}{c^2}\right)\delta_{ij}+{\cal O}\left(\frac{1}{c^4}\right)\;,
\end{align}
from which we can read off the parameters $\alpha_1$ and $\alpha_2$
\begin{align}
\alpha_1&=\frac{4(\alpha-2\beta)}{\beta-1}\;,\\
\alpha_2&=\eta_{0,1}\frac{2(1-\alpha+3\beta+2\lambda)(2+\beta+3\lambda)}{3(\alpha-2)(\beta+\lambda)}\nonumber \\ &+ \frac{(\alpha-2\beta)[-\beta(3+\beta+3\lambda)-\lambda+\alpha(1+\beta+2\lambda)]}{(\alpha-2)(\beta-1)(\beta+\lambda)}\;.
\end{align}

\bibliographystyle{utphys}
\bibliography{bh_HG}

\end{document}